\documentclass[
    ,final            % use final for the camera ready runs
  ]
  {aipproc}

\layoutstyle{6x9}

\begin{document}

\title{Measuring the low mass end of the M$_{\bullet} - \sigma$ relation}


\author{Davor Krajnovi\'c}{
  address={Sub-department of Astrophysics, University of Oxford, Denys Wilkinson Building, Keble Road, OX1 3RH, UK}
  ,altaddress={European Southern Observatory, Karl-Schwarzschild-Str. 2, 85748 Garching bei M\"unchen, Germany}
}

\author{Richard M. McDermid}{
  address={Gemini Observatory, Northern Operations Center, 670 N. A'ohoku Place,Hilo, Hawaii, 96720, USA}
}

\author{Michele Cappellari}{
  address={Sub-department of Astrophysics, University of Oxford, Denys Wilkinson Building, Keble Road, OX1 3RH, UK}
  %,altaddress={<author1 address>} % additional visiting address
}

\author{Roger L.\ Davies}{
   address={Sub-department of Astrophysics, University of Oxford, Denys Wilkinson Building, Keble Road, OX1 3RH, UK}
}

\begin{abstract}
We show that high quality laser guide star (LGS) adaptive optics (AO) observations of nearby early-type galaxies are possible when the tip-tilt correction is done by guiding on nuclei while the focus compensation due to the changing distance to the sodium layer is made 'open loop'. We achieve corrections such that 40\% of flux comes from R$<$0.2 arcsec. To measure a black hole mass (M$_{\bullet}$) one needs integral field observations of both high spatial resolution and large field of view. With these data it is possible to determine the lower limit to M$_{\bullet}$ even if the spatial resolution of the observations are up to a few times larger than the sphere of influence of the black hole. 
\end{abstract}

\maketitle

%%%%%%%%%%%%%%%%%%%%%%%%%%%%%%%%%%%%%%%%%%%%
%% MAINMATTER
%%%%%%%%%%%%%%%%%%%%%%%%%%%%%%%%%%%%%%%%%%%%

\section{Introduction}

The correlation between the masses of supermassive black holes (SMBH) with the global characteristics of their host galaxies \citep[e.g.][]{2000ApJ...539L...9F,2000ApJ...539L..13G} is crucial for our interpretation of galaxy formation and evolution \cite[e.g.][]{1998A&A...331L...1S,2006ApJS..163....1H}.  M$_{\bullet}$ is usually determined using dynamical tracers such as a disk of gas clouds in Keplerian rotation around the SMBH or the line-of-sight velocity distribution (LOSVD) of stars.  Both methods require observations of high spatial resolution since the dynamical models have to be constrained by resolved kinematics which probe the sphere of gravitational influence of the SMBH (R$_{sph}$), defined as the distance from SMBH at which the potential of the galaxy and SMBH are approximately equal. This spatial scale, usually defined as R$_{sph}= GM_{\bullet}/\sigma^2$, where $\sigma$ is the velocity dispersion of the galaxy, is typically significantly less than an arcsec in apparent size, even for nearby galaxies. 

The influence of the SMBH, however, will be felt at larger radii than R$_{shp}$, albeit  to a lessening degree. Here we show that it is possible to constrain the M$_{\bullet}$ when R$_{shp}$ is up to  3 times smaller than the spatial resolution of the observations if the observation consists of {\it (i)} low (arcsec) spatial resolution integral field data covering the galaxy out to about 1 effective radius required to determine the overall dynamical mass-to-light ratio of the system, {\it (ii)} high (sub-arcsec) resolution integral field data probing the stellar kinematics in the vicinity of the SMBH and have {\it (iii)} sufficient signal-to-noise ratio (S/N) to extract higher order moments of the LOSVD.

In this report we summarise the results of \cite{2009MNRAS.399.1839K} which present the usage of a new observational method with LGS AO system especially suitable for determination of M$_{\bullet}$ in the nuclei of nearby low mass early-type galaxies. For determination of M$_{\bullet}$ in massive early-type galaxies using similar method, but non-AO observations see the contribution of \cite{Cappellari2010} in this proceedings.

\section{Measuring small M$_{\bullet}$ from ground}

The main prohibitive issue with AO observations is that, even with LGS capabilities, a natural guide star is required for an optimal correction of atmospheric aberrations and there are only a handful of galaxies with a near-resolvable R$_{sph}$ that have a suitable guide star. This limitation has prompted innovative use of AO techniques, such as neglecting altogether the low-order corrections provided by the natural tip-tilt  \citep{2008Msngr.131....7D}. In this study, we employ a novel method developed at Gemini Observatory for Altair LGS AO system.

\subsection{Open loop focus model of LGS}

To avoid the issue of suitable guide stars, it is, in principle, also possible to use the nucleus of the galaxy as a natural guide source for the system, provided it is sufficiently bright and compact (drop of $\geq 1$ magnitude within the central $\sim 1$ arcsec for Altair). The central light profiles of early-type galaxies show a general change from steeply rising `cusp' profiles at lower masses, towards a flatter central profile that can define a central `core' region \citep[e.g.][]{1997AJ....114.1771F}. In order to test what correction of the PSF could be expected for typical early-type galaxies, we chose NGC524 and NGC2549 which both satisfy the basic Altair requirement for nuclear guiding. Specifically, NGC524 has a core-like light profile, while NGC2549 a cusp-like profile at the HST resolution. In this proceeding we will consider the observations of NGC2549 only, the smaller of the two galaxies. 

After initial observations were attempted, it became clear that the chosen nuclei, while suitable for tip-tilt correction, are too faint to be used for constraining the focus. Gemini staff implemented a procedure by which the focus correction during the science integration is controlled by a geometric function that takes into account the change in the distance to the sodium layer as the telescope position changes. In this, so-called `open-loop' focus model, the only time-dependent parameter is the altitude of the sodium layer, which is determined immediately before observing the galaxy by `tuning' the LGS AO system using a nearby bright star. When all the control loops of the system (tip-tilt, focus, and LGS) have converged with this reference source, the loops are opened and the science target is acquired. The tip-tilt and LGS control loops are then closed, using the galaxy nucleus and laser beacon respectively as reference sources. The focus loop is left open, being passively controlled by the open-loop model. After approximately one hour of science observations, the bright reference star is re-observed, so that the degradation of the PSF can be estimated (our tests show minimal degradation) and the LGS AO system can be re-optimised for further observations. 

\begin{figure}
  \includegraphics[height=.25\textheight]{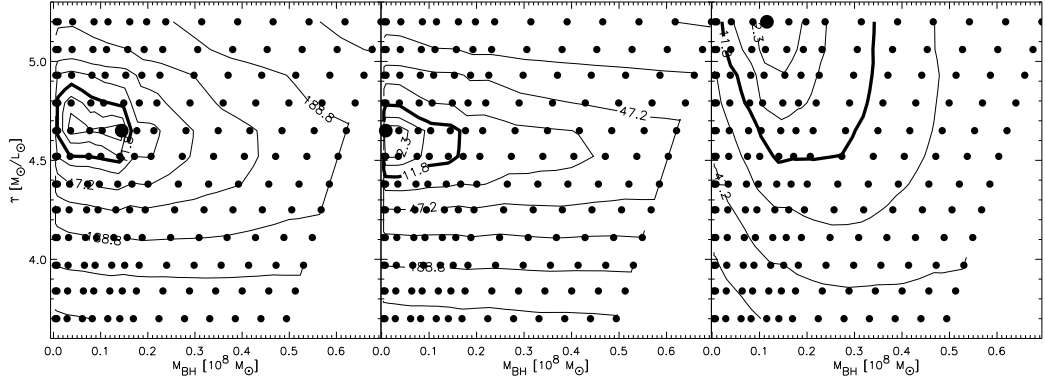}
  \caption{\label{f:grid} Schwarzschild dynamical models and the determination of the best fitting parameters for NGC2549. Each symbol is a dynamical model. The agreement between the data and the models are described by overploted contours $\Delta  \chi^2$ contours showing 1, 2 and 3$\sigma$ levels for two parameters. Further contours are spaced by a factor of 2. Large symbol on each panel marks the best fitting models. {\bf From left to right} panels show grids of models constrained by both SAURON and NIFS kinematics, models constrained by SAURON only kinematics and models constrained only using NIFS data.}
\end{figure}

We estimated the PSF of the science observations by convolving (with a double Gaussian) and rebining (to the pixel size of our observations) an HST/WFPC2 image. This image is compared with the reconstructed image of our observations, and the parameters of the PSF are varied until the best matching double-Gaussian is found. The final PSF of our LGS AO observations was: 0.17 and 0.80 arcsecs (FWHM) for the the narrow and broad Gaussian components and intensities of 0.53 and 0.47, respectively.

\begin{figure}
  \includegraphics[height=.2\textheight]{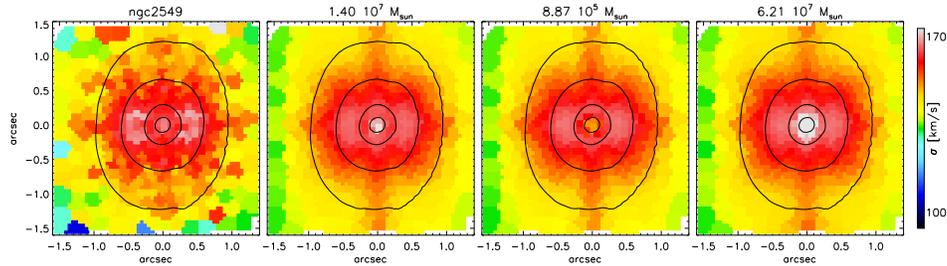}
  \caption{\label{f:maps} Comparison between data and model velocity dispersion maps for NGC2549. From left to right: $\sigma$ (symmetrised) and model prediction maps for different M$_{\bullet}$ (at the best fitting M/L). The best fitting model is shown in the map adjacent to the observed $\sigma$ map, followed by a model with a too small M$_{\bullet}$ and a model with too large M$_{\bullet}$. The models were constrained using both SAURON and NIFS data. Note the colour change in the central few pixles which are a consequence of change in M$_{\bullet}$.}
\end{figure}

\subsection{Determining M$_{\bullet}$ when the resolution is smaller than R$_{sph}$}

We observed NGC2549 with NIFS IFU centered on CO absorption features starting at 2.29 microns. We spatially binned the data using Voronoi method \cite{2003MNRAS.342..345C} and achieved the SN of $\sim60$, while within the central arcsec the bins are no larger than $0.1 \times 0.1$ arcsecs. We extracted stellar kinematics using the pPXF method \cite{2004PASP..116..138C}. Large scale kinematic data were obtained from previous observations with SAURON IFU \cite{2001MNRAS.326...23B} and were previously presented in  \cite{2004MNRAS.352..721E}. We constructed orbit-based Schwarzschild dynamical models following the method presented in \cite{2006MNRAS.366.1126C}.

Fig.~\ref{f:grid} shows grids of Schwarzschild dynamical models constrained using three different kinematic data sets: combined SAURON and NIFS data, SAURON data only and NIFS data only. The best fit model using SAURON and NIFS kinematics gives M$_{\bullet}$=($1.4^{+0.2}_{-1.3} )\times 10^7$ M$_{sun}$. It is evident that when using only SAURON data it is not possible to determine the lower limit of the M$_{\bullet}$, simply because the resolution of the SAURON data is 1.7 arcsec, while R$_{sph}$=0.05 arcsec, using the above best fit model. Using the NIFS dataset only it seems possible to determine even the lower limit, in spite of 0.17 arcsec resolution, but the uncertainty on M$_{\bullet}$ is very large because, due to the small field-of-view, NIFS data alone are not able to constrain the total orbital distribution and give the right mass-to-light ratio. These are, however, well constrained using SAURON large field-of-view (FoV) observations. Combining the two data sets, the high resolution and large FoV, one can constrain Schwarzschild models and determine the M$_{\bullet}$. 

Fig.~\ref{f:maps} illustrates that our NIFS observations, although at about 3 times lower resolution than the estimated R$_{sph}$, are able to capture the change in the kinematics influenced by the SMBH. We illustrate this on the velocity dispersion: the $\sigma$ map of the best fit model is more similar to the observed data than $\sigma$ maps of models with a too small or a too big SMBH. To see this effect it was however, necessary to have high quality integral field data both at high resolution (NIFS) and with large FoV (SAURON).

%\section{Summary}
%%%%%%%%%%%%%%%%%%%%%%%%%%%%%%%%%%%%%%%%%%%%
%% Sample figure:
%%
%% The option [height=...] scales the picture to the given height,
%% without it it would be printed at its nominal size
%%%%%%%%%%%%%%%%%%%%%%%%%%%%%%%%%%%%%%%%%%%%

%%%%%%%%%%%%%%%%%%%%%%%%%%%%%%%%%%%%%%%%%%%%%%%%
%% BACKMATTER
%%%%%%%%%%%%%%%%%%%%%%%%%%%%%%%%%%%%%%%%%%%%%%%%

%\begin{theacknowledgments}
%\end{theacknowledgments}



\bibliographystyle{aipproc}   % if natbib is available

%\bibliography{sample}
%\bibliography{/Users/dkrajnov/WORK/papers/sci/refs.bib}

%  \typeout{******************************************}
%  \typeout{** Please run "bibtex \jobname" to optain}
%  \typeout{** the bibliography and then re-run LaTeX}
%  \typeout{** twice to fix the references!}
%  \typeout{******************************************}

%\end{document}

%%%%%%%%%%%%%%%%%%%%%%%%%%%%%%%%%%%%%%%%%%%
%% The following lines show an example how to produce a bibliography
%% without the help of the BibTeX program. This could be used instead
%% of the above.
%%%%%%%%%%%%%%%%%%%%%%%%%%%%%%%%%%%%%%%%%%%

\end{document}